# Enhancement of vortex liquid phase and reentrant behavior in NiBi$_3$ single crystals

V. Rollano[1*], M. C. de Ory[1,2], A. Gomez[2], E. M. Gonzalez[1,3], Z. Pribulová[4], M. Marcin[4], P. Samuely[4], G. Sanchez-Santolino[5&], A. Torres-Pardo[6], F. Mompean[5], M. García-Hernández[5], I. Guillamón[7], H. Suderow[7], M. Menghini[1] and J. L. Vicent[1,3]

[1]IMDEA-Nanociencia, Cantoblanco, 28049 Madrid, Spain

[2]Centro de Astrobiología (CAB), CSIC-INTA, 28850 Torrejón de Ardoz, Madrid, Spain

[3]Departamento Física de Materiales, Universidad Complutense de Madrid, 28040 Madrid, Spain

[4] Centre of Low Temperature Physics, Institute of Experimental Physics SAS, 040 01 Košice, Slovakia

[5]Instituto de Ciencia de Materiales de Madrid (ICMM-CSIC), 28049 Madrid, Spain

[6]Departamento de Química Inorgánica, Universidad Complutense de Madrid, 28040 Madrid, Spain

[7] Laboratorio de Bajas Temperaturas y Altos Campos Magnéticos, Unidad Asociada UAM, CSIC, Departamento de Física de la Materia Condensada, Instituto Nicolás Cabrera and Condensed Matter Physics Center, Universidad Autónoma de Madrid, 28049 Madrid, Spain

*Present address: Shanghai Branch, CAS Center for Excellence in Quantum Information and Quantum Physics, University of Science and Technology of China, Shanghai 201315, China and Instituto de Nanociencia y Materiales de Aragón (CSIC-Aragón), 50009 Zaragoza, Spain

&Present address: Departamento de Física de Materiales & Instituto Pluridisciplinar, Universidad Complutense de Madrid, 28040 Madrid, Spain

E-mail: mariela.menghini@imdea.org

We investigated the vortex phase diagram of needle shaped high quality NiBi$_3$ single crystals by transport measurements. The current is applied along the crystalline *b*-axis of this intermetallic quasi-1D BCS superconductor. The single crystals show a Ginzburg-Levanchuk (G$_i$) parameter few orders of magnitude larger than other low T$_c$ BCS superconductors. Vortex phase diagram, critical currents and pinning forces have been extracted from the experimental data. The main findings are: 1) Enhancement of the vortex liquid phase in comparison with low T$_c$ superconductors, 2) reentrance of the liquid phase at low fields and 3) deviation of the pinning force vs field from the usual pinning mechanisms. The interplay between weak pinning, due to quenched disorder, and the quasi-1D character of the material could be a hint to explain the lack of a single pinning mechanism.

## INTRODUCTION

Superconductivity in compounds containing magnetic elements is a topic which has called the attention of many researchers. The literature on this subject span from theoretical to experimental works. Within this family, binary intermetallic compounds are interesting as they possess a simple crystalline structure and relatively high critical temperature. NiBi$_3$ is an intermetallic compound that shows BCS-like superconductivity at temperatures below ~ 4 K (1-3). The crystalline structure of NiBi$_3$ is orthorhombic with lattice parameter ratio $a/b$ ($c/b$) of ~ 2 (2.8) (4, 5). This quasi-1D crystalline structure gives electronic mass ratios $m_a/m_b \cong m_c/m_b \cong 5 - 7$ (1, 6) corresponding to an anisotropy $\varepsilon = (m_b/m_{a,c})^{1/2} \cong 0.37 - 0.44$ and leads to the formation of superconducting vortices elongated along the *b*-axis when the magnetic field is applied in a direction perpendicular to that crystalline axis (4). There is still some controversy on the possible interplay between superconductivity and magnetism in NiBi$_3$. On one hand, signatures of bulk and surface intrinsic magnetism have been reported based on magnetization (7) and electron spin resonance (1), respectively. On the other hand, density functional theory calculations show that ferromagnetic interactions seem to not play a significant role in this compound (8). More recently, a coexistence of superconductivity and magnetism, likely due to inclusions of Ni clusters, has been observed (5).

It is known that the presence of fluctuations and defects in superconductors can transform the simple Abrikosov vortex lattice in a variety of phases including glassy solid phases (9-14), liquid phases (15) or even intermediate ones like the vortex slush (16, 17). In short, fluctuations and disorder give rise to a complex phase diagram for the mixed state in type-II superconductors. The transition from a vortex solid phase to a liquid one has been of intense research both from the experimental and theoretical points of view (16-28).

The vortex lattice melting line has been calculated (16, 17) considering a nonlocal elasticity theory based on the phenomenological Lindemann criterion. According to this criterion, the melting of the lattice takes place when the root mean square (rms) of the thermal vortex displacement fluctuations is as large as a certain fraction, $c_L$, of the vortex lattice parameter. The melting field line, derived from these theoretical calculations, is estimated as the following expression (29)

$$B_{sl}(T) \approx H_{c2}(0) \frac{4\theta^2}{(1+\sqrt{1+4\theta T_S/T})^2} \quad (1)$$

with $\theta = c_L^2 \sqrt{(\beta_m/G_i)} \left(\frac{T_c}{T} - 1\right)$, $T_S = T_c c_L^2 \sqrt{\beta_m/G_i}$, $T_c$ is the critical temperature of the superconductor, $H_{c2}(0)$ is the upper critical field at *T = 0* K, $G_i$ is the Ginzburg-Levanyuk number and $\beta_m$ is a constant $\approx$ 5.6. Although this theoretical study was developed to explain the effect of thermal fluctuations in the vortex lattice melting in oxide superconductors, the resulting $B_{sl}(T)$ adjusts reasonably well to the experimental results found in other types of superconductors. For example, this estimation fits the liquid-solid transition of the borocarbide superconductor

YNi$_2$B$_2$C determined from magnetoresistance measurements for fields lower than 1T and considering $c_L$ = 0.02 (30). The irreversibility lines, extracted from magnetization measurements, in Nb, Nb$_3$Sn and Nb-Ti alloys have been also associated with the vortex lattice melting model mentioned above (31, 32). In these last cases, the Lindemann numbers are larger (0.04-0.1) than the case of the borocarbide compound but still smaller than the ones found for the cuprates (0.1-0.4) (29).

The fraction of the *H-T* phase diagram occupied by the vortex liquid phase depends on $G_i$ which measures the importance of thermal fluctuations compared to fluctuations of the order parameter (29, 33). For high $T_c$ materials this number is relatively large ($G_i \sim 10^{-3}$), leading to a vortex liquid phase that occupies a large portion of the phase diagram. On the other hand, the extremely small contribution of thermal fluctuations in conventional superconductors ($G_i < 10^{-9}$) narrows the liquid phase area closely to the upper critical field making it, in most of the cases, experimentally inaccessible (32). Finally, there is a range of superconductors such as intermetallic compounds, borocarbides and pnictides with intermediate $G_i$ numbers ($10^{-7}$-$10^{-5}$) (30, 31).

The literature related to superconductivity in NiBi$_3$ is mostly limited to the study of the anisotropic superconducting properties and the role of magnetism in single crystals of this compound (1-8). Artificial layered structures of Ni and Bi have been used to study the coexistence of superconductivity and magnetism (34). In thermally evaporated Ni/Bi bilayers, NiBi$_3$ forms spontaneously by a reaction-diffusion process of Bi (35). Besides, it was proven in the same work that superconductivity in NiBi$_3$ is quite robust against ion irradiation. Very recently, the growth of NiBi$_3$ thin films has been reported by a co-deposition method where the rate of Bi deposition seems to have a large influence in the superconducting and magnetic properties of the films (36). Ni/Bi bilayers and NiBi$_3$ thin films can provide a playground to study superconducting proximity effect in a clean superconductor-ferromagnet interface and the coexistence of superconductivity and ferromagnetism with controlled impurities.

To our knowledge, there are no reports on vortex dynamics and vortex pinning properties in NiBi$_3$. A detailed description of the vortex phase diagram in NiBi$_3$ is relevant for the study of proximity effects, coexistence of superconductivity with magnetism and possible applications of this material in high radiation environments. Furthermore, it can also enrich the understanding of vortex dynamics in quasi-1D BCS superconductors.

In this work, we focus on the study of the vortex dynamics and critical currents in NiBi$_3$ single crystals by analyzing voltage-current *(V-I)* characteristics in different regions of the *H-T* phase diagram. Our results are consistent with a vortex solid-liquid transition line and with a reentrant behavior for fields below ~ 6 mT. We found that the transition line, for fields larger than the reentrant field, follows the prediction of the elastic theory based on the Lindemann criterion with $c_L$ ~ 0.03. In the low-field reentrant region the critical current and pinning force decrease when decreasing the magnetic field for temperatures close to the $H_{c2}$ line. Interestingly, the behavior of the pinning forces as a function of the reduced field does not follow what is expected for traditional pinning mechanisms reported in the literature. The observed vortex dynamics and critical current behavior at very low magnetic fields could be related with the combination of the quasi-1D character of NiBi$_3$ and the low bulk pinning present in these single crystals.

**EXPERIMENTAL DETAILS**

Single crystals of NiBi$_3$ were grown by the high temperature flux method as reported by some of us in Ref. (5). The crystals grown by this method have an orthorhombic structure and needle-like shape (1).

Scanning transmission electron microscopy (STEM) and Energy-dispersive X-ray spectroscopy (EDS) measurements were performed on an aberration-corrected JEOL JEM ARM-200 cF electron microscope operated at 120 kV, equipped with a cold field emission gun and an Oxford Instruments EDS spectrometer. Atomic resolution high angle annular dark field (HAADF) and annular bright field (ABF) images of NiBi$_3$ needles reveal the high crystalline quality of the samples confirming that the *b* crystalline axis is oriented along the length of the needles as previously reported (1), see Fig. 1. In the ABF image it is possible to identify the Ni columns in some locations (not visible in the HAADF image) as weak contrast changes in between the more contrasting pairs of Bi atoms. Energy-dispersive X-ray spectroscopy EDS quantification results from the spectrum in Fig. 1(c) show that the relative Ni and Bi composition in the studied single-crystals is homogeneous within the sample deviating only ~ 5% from the expected concentration in stoichiometric NiBi$_3$. These results evidence the high structural quality of the NiBi$_3$ single crystals.

Magneto-transport measurements were performed by the usual four probe dc technique with the applied current parallel to the long direction of the needles using a commercial $^4$He cryostat equipped with a 9 T superconducting solenoid, and a variable temperature insert. The crystals used for the transport measurements had approximate dimensions of about 130 μm wide and 150 μm thick. Four in-line electrical contacts were made along the length of the needles with voltage contacts separation of 450 μm. From the resistance vs temperature measurements, we obtained a zero-field transition temperature $T_c$ = 4.13 K with a transition width of around 100 mK and a residual resistance ratio (R$_{300K}$/R$_{4.3K}$) RRR=12.9. SQUID magnetization loops below $T_c$ performed on a single crystal from the same batch show perfect diamagnetism as expected for the superconducting phase. These results are comparable with previous work in similar crystals (5). The calculated with previous work in similar crystals (5). The calculated mean free path of the NiBi$_3$ single crystals is $l = 29$ nm.

The V-I characteristics (either as a function of temperature at fixed magnetic field or vice-versa) are a powerful tool to investigate vortex dynamics and vortex phases in the mixed state of a superconductor (37). *V-I* curves were measured in NiBi$_3$ single crystals using a temperature step of 10 mK with a temperature stability of ~ 1 mK. The critical current was defined using a voltage criterion of 50 nV. The superconducting penetration depth, *λ*, of NiBi$_3$ was determined using local Hall-probe magnetometry in a single crystal from the same batch as the ones used in the magneto-transport experiments. The sample with dimensions

width = 130 µm, length = 725 µm and thickness = 80 µm was placed on top of a high-sensitivity (~1kΩ/T) Hall-probe array powered by 10 µA constant current. The array consisted of a row of 5 Hall sensors aligned in the direction perpendicular to the length of the NiBi$_3$ needle. The magnetic field was applied in the same direction as in the transport measurements (see insert in Fig. 3(d)). Probes with 10 x 10 µm$^2$ active area and pitch of 25 µm have been used to determine the field distribution over a length span of ~100 µm across the sample. The resulting magnetic field profile of the sample was dome-shaped, characteristic of low pinning. In such a case the first vortex settles close to the sample center. Therefore, a probe close to the center was selected to determine the penetration field $H_p$ - a field when first complete vortex appears in the sample - as a function of temperature from 4 K down to 0.4 K. Lower critical magnetic field $H_{c1}$ can be calculated from the penetration field using an expression $\mu_0 H_{c1} = \frac{\mu_0 H_p}{\tanh\sqrt{\alpha \frac{d}{2w}}}$, where d and 2w is the thickness and width of the sample and coefficient $\alpha$ is 0.67 for disk-like crystal and 0.36 for an infinite stripe (38). For our sample, we considered $\alpha$ being average of the two extreme cases. The penetration depth λ was then calculated using the relation $\mu_0 H_{c1} = \frac{\phi_0}{4\pi\lambda^2}[\ln\kappa + a(\kappa)]$, where κ is Ginzburg-Landau parameter and $a(\kappa) = 0.5 + \frac{1+\ln 2}{2\kappa - \sqrt{2} + 2}$ (39). The resulting temperature dependence of $H_{c1}$ is depicted in Fig.2, in the zero temperature limit we find $\lambda(0) \sim 190 \text{ nm} \pm 15 \text{ nm}$. More details about the method can be found in Ref. (40).

**RESULTS**

One of the main differences between the vortex solid and liquid phase is the existence of infinite barriers in thermodynamic equilibrium in the solid and their absence in the liquid phase (29). In fact, Strachan *et al.* (41) have shown that a change in curvature in the derivative *d(log V)/d(log I)* vs current can be associated to a transition between a vortex solid and a vortex liquid phase. More specifically, under this criterion the transition temperature, $T_{sl}$, is defined as the temperature for which the derivative changes concavity for curves measured at a fixed external magnetic field. From the *V-I* measurements in NiBi$_3$ as a function of temperature for different magnetic fields, we analyzed the change in behavior of the *V-I* characteristics derivatives (*d(log V)/d(log I)*) in a similar way as reported in (41) and defined the error for the extracted $T_{sl}$ as the temperature step between consecutive *V-I* curves.

Figures 3(a) and (b) show examples of *V-I* curves in log-log scale measured at different temperatures, with the current flowing parallel to the crystalline *b*-axis (see Fig. 1(a) and insert of Fig. 3(d)), for applied magnetic fields of 20 mT and 1 mT, respectively. The high temperature region is characterized by a linear behavior in log-log scale. As temperature decreases, non-linearities in the *V-I* characteristics develop. Similar trends have been observed in the whole range of fields studied in this work and a in *V-I* measurements done at a fixed temperature and varying the magnetic field. Figures 3(c) and (d) depict the derivatives, *d(log V)/d(log I)* vs current, of the curves shown in Figs. 3(a) and (b), respectively. The transition from the linear to non-linear response provides the temperature $T_{sl}$ defined as the temperature in which a change in the concavity of the derivatives is observed (41, 42). This transition temperature is estimated to be 3.915 K for 20 mT and 4.005 K for 1 mT.

This previous analysis done for the curves shown in Fig. 3 was repeated for a large set of applied magnetic fields, allowing the determination of the transition line, $H_{sl}(T)$, while the upper critical field, $H_{c2}(T)$, was obtained from the R vs T measurements at constant magnetic field using the $0.5\,R_N$ criterion ($R_N$ is the normal state resistance just above the transition) to define the transition temperature. The superconducting coherence length has been extracted from the slope of $H_{c2}(T)$ close to $T_c$ and using the WHH approach (43) giving a value of $\xi(0) \sim 23.4$ nm.

In Fig. 4(a) the obtained values for $H_{sl}(T)$ together with the upper critical field, $H_{c2}(T)$, are depicted with orange circles and black squares, respectively. The resulting H-T phase diagram for $NiBi_3$ single crystals shows that $H_{sl}(T)$ lies significantly below the $H_{c2}(T)$ line. Considering the obtained $\xi(0)$ and $\lambda(0)$, and taking an average value for the anisotropy $\varepsilon^2 = 0.17$ (1, 6) we find that $G_i = 7.23 \cdot 10^{-8}$ for our samples. Interestingly, $G_i$ is few orders of magnitude larger than the typical values for conventional low-$T_c$ materials but still significantly smaller than for high $T_c$ superconductors. With this value for $G_i$ and $H_{c2}(0) = 595.2$ mT extrapolated from a linear fit to the $H_{c2}(T)$, we find that the experimental $H_{sl}(T)$ line (for H > 6 mT) follows Eq. (1) quite well with a single fitting parameter $c_L = 0.032 \pm 0.001$. The result of the fitting is shown by the red line in the insert of Fig. 4(a). The observation of a vortex liquid-solid transition line clearly below the $H_{c2}(T)$ line is surprising given the low $T_c$ and relatively large $\xi$ of this compound. Besides, for fields below 6 mT there is a reentrance of $H_{sl}(T)$. This low-field reentrance is also evident when analyzing V-I characteristics measured at fixed temperatures as a function of the magnetic field. As an example, a series of d(log V)/d(log I) vs current at 4.03 K for different magnetic fields is shown in Fig. 4(b). The non-divergent d(log V)/d(log I) at low currents for H = 0.5 and 0.75 mT correspond to a regime of finite resistance when I → 0 while at H = 3.5 and 4.5 mT the upward trend of the derivatives indicates a transition to a non-linear regime (solid phase). At 10 mT, a linear regime (evident as a constant d(log V)/d(log I)) is observed.

change in the derivative of V-I curves as described in the text indicating the vortex solid-liquid transition (the dashed line is a guide to the eye). Insert: The red line shows the solid-liquid transition line obtained from Eq. (1). The green diamonds correspond to the $H_{c1}$ line from Fig. 2. (b) d(log V)/d(log I) vs current at T = 4.03 K for different magnetic fields as indicated with coloured crosses in (a). (c) Vortex lattice parameter at the solid-liquid transition, $a_0(H_{sl})$, and $\lambda_{BCS}$ vs temperature. The reentrance of the liquid phase at H = 6 mT takes place when $a_0 \sim \lambda$.

Finally, it is important to analyze the behavior of the critical currents to address the influence of quenched disorder on the vortex matter in $NiBi_3$ (37). On one hand, as shown in Fig. 5, for fields H > 10 mT the critical current density, $J_c$, decreases as the temperature increases while it decreases for larger fields as expected. The reduced temperature $t = T/T_{c2}(H)$ is defined as the temperature normalized by the transition temperature at each corresponding field H, $T_{c2}(H)$. On the other hand, an unusual behavior is observed in the very low field regime, see Fig. 6(a). For fields down to ~3 mT, $J_c$ increases while for H < 2 mT it decreases as the field is reduced. This implies that at a fixed reduced temperature the critical current has a dome-shaped behavior as a function of magnetic field as can be clearly seen in Fig. 6(b). From the critical currents extracted from the V-I characteristics, we have also calculated the pinning force ($F_p = J_c \times H$) as a function of t at different applied fields, see Fig. 6(c). In the low field region, the slope of $F_p$ with temperature significantly decreases as the field is reduced.

In summary, we have made a complete characterization of the transport properties of NiBi$_3$ single crystals as a function of magnetic field and temperature. We observe changes in the dynamic vortex behavior from highly non-linear I-V curves to a large highly dynamic (or liquid-like) regime close to the $H_{c2}$ line with a reentrance to lower temperatures at very low magnetic fields.

**DISCUSSION**

A reentrance of the vortex melting line $H_{sl}(T)$ has been predicted for fields low enough such that the vortex separation is larger than the penetration length [29, 44]. This re-entrant behavior implies a reduction and a subsequent increase of vortex mobility when increasing the magnetic field. We have calculated the vortex lattice parameter $a_0(H_{sl})$ for each ($H_{sl}$, T$_{sl}$) point of the NiBi$_3$ phase diagram (Fig. 4(a)). In Fig. 4(c) we compare $a_0(H_{sl})$ at the corresponding transition temperature, $T_{sl}$, with the evolution of $\lambda(T)$ assuming a BCS behavior. We observe that the reentrance of the $H_{sl}$ line, signalled by the turning point of $a_0$, takes place when $a_0(H_{sl}) \sim \lambda$. This result indicates that the observed reentrance of the $H_{sl}$ line is a consequence of the vortex lattice softening at low fields. When the field is further decreased and $a_0$ increases, the vortex-vortex interaction is weaker and consequently the melting will take place at lower temperatures. In our crystals, in this regime there is a finite small critical current in the liquid phase. The presence of disorder can cause a shift of the metling line (22) and could also induce a transition to a pinned-liquid (15) or entangled liquid phase (29). The identification of the resulting vortex phase in the low reentrant field region cannot be univocally determined by solely transport measurements and goes beyond the scope of the present paper.

The relevance of quenched disorder in the vortex matter can be evaluated by calculating the ratio between the critical current and the depairing current $J_c(T,H)/J_0(T)$ where $J_0(T) = c\Phi_0/12\sqrt{3}\pi^2\lambda^2(T)\xi(T)$ (22, 29, 45). We have calculated this ratio at T = 3.5K (that corresponds to $T/T_c$ ~ 0.85) and H = 10 mT in the studied NiBi$_3$ single crystals where $J_c$ ~ 1 10$^6$ A/m$^2$ (see Fig. 5). Using the value of $\lambda(0)$ extracted from the local magnetization measurements and the BCS expression for $\lambda(T)$ we obtain, at this field and temperature, $\frac{J_c}{J_0}$ ~ 5 10$^{-5}$ (for higher fields $J_c/J_0$ decreases as the critical current decreases with field). Similar critical current ratios have been found in different type of materials such as pyrochlore (45), high-$T_c$ (46) and topological superconductors (47). For the pyrochlore and topological superconductors these values were found for reduced temperatures $\frac{T}{T_c} \sim 0.7 - 0.8$. In the case of pyrochlore RbOs$_2$O$_6$ single crystals, it has been shown that the low bulk pinning, and consequently low value of $J_c$ with respect to $J_0$, favors the existence of a wide reversible vortex region below $H_{c2}$ (45). These low values for $J_c/J_0$ in NiBi$_3$ could be an indication that pinning due to quenched disorder is very weak in the low field region.

It is important to note that the observed maximum in $J_c$ (see Fig. 6(b)) does not seem to correspond to the ubiquitous peak effect or second magnetization peak observed in many superconductors associated to a weakening of the vortex lattice as the $H_{c2}$ line is approached or a transition from elastic to plastic behavior (48-52). In the present case, there is no peak in the critical current vs temperature, as expected for the reported peak effect, close to the $H_{c2}$ line in the range of fields and temperatures studied in this work (~2.5 K < $T$ < $T_c$, 0 < $H$ ≤ 200 mT).

The behavior of the pinning force as a function of magnetic field can provide information related to the relevant pinning mechanism in superconductors (53-56). If a single pinning mechanism dominates the critical current response, the pinning force shows a scaling behavior as a function of a reduced field, $b$, as $F_p \propto b^p(1-b)^q$ known as the Dew-Hughes model (57). For conventional type-II superconductors the reduced field $b$ is normally defined as $H/H_{c2}$ while for high-$T_c$ superconductors the same scaling law was observed when normalizing by the irreversibility field, $H_{irr}$ (53, 54, 56) (i.e. the field at which the critical current vanishes) although there is some debate in the literature on how to correctly estimate this field (58). In the case of NiBi$_3$ single crystals, the normalized pinning force as a function of $b$ (see Fig. 6(d)) in the range of fields below 10 mT follows a single peaked function for temperature between 3.93 K and 4.04 K corresponding to the reentrant region in the phase diagram (15). The exponents p=2.1 and q=7.7 that fit the experimental data points are larger than usually reported for pinning by only point disorder (p=1 and q=2) or magnetic pinning (p=0.5 and q=1) (57, 59). An improved version of the Dew-Hughes model (57) considering different types of pinning centres acting simultaneously has been used in Ref. (60). In this model, the pinning force can be written as $F_p = a_1 f_1 + a_2 f_2 + ... + a_i f_i + \cdots a_6 f_6$ where $f_i = h^{p_i}(1-h)^{q_i}$ and $a_i \geq 0$. We have used this formula to fit our data but we have obtained negative values for some of the $a_i$ coefficients, which has no physical sense. It is interesting to note that the Dew-Hughes model has been extensively used to explain the pinning mechanism in a large variety of superconducting materials. Our results indicate that a more complex pinning mechanism can play a role in the low field region in NiBi$_3$ single crystals. It is reasonable to argue that the possible interplay between magnetism and superconductivity (61) combined with the quasi-1D character of the material could have an influence in the pinning mechanisms relevant in this range of magnetic fields where a reentrance of the vortex liquid phase is observed. Low temperature local probe techniques could shed light on this matter, although this goes beyond the scope of the present work.

**CONCLUSIONS**

Magneto-transport measurements in NiBi$_3$ single crystals have revealed a vortex solid-liquid transition near the *H$_{c2}$(T)* line and an enhancement of the vortex liquid phase at low fields. This reentrance, revealed by a change in the behavior in the derivative of the *V-I* curves, develops at fields lower than 6 mT where the lattice parameter is of the order of or larger than the penetration length. The low value of the critical current compared to the depairing current indicates that quenched disorder in our single crystals is weak which can favor the observed enhancement of the liquid phase in a superconductor with a relatively low $G_i$. In the low field region, below the reentrance of the vortex solid-liquid transition line, a decrease in critical current and in pinning force is observed as the field decreases. The normalized pinning force scales with the reduced magnetic field showing a single-peaked function as reported for a large variety of superconductors. However, the obtained scaling exponents indicate that in this field region vortex pinning is not determined by a single pinning mechanism. The maximum in critical current as a function of field could be related to the reentrance of the liquid phase and the more complex pinning mechanism that plays a role in this field region. The observed low field *H-T* phase diagram suggests that there could be interesting vortex phases to be analyzed by local probe techniques in this field range.

**Acknowledgements**

The work has been supported by Spanish MICINN grants FIS2016-76058 (AEI/FEDER, UE), EU COST- CA16218 (Nanocohybri). IMDEA Nanociencia acknowledges support from the 'Severo Ochoa' Programme for Centres of Excellence in R&D (MICINN, Grants SEV-2016-0686 and CEX2020-001039-S). EMG and MM acknowledge support from PID2021-122980OB-C52. MCO and AG acknowledge financial support from Spanish Ministry of Science and Innovation under Grant PID2019-105552RB-C4-1 and ONR-Global under support under Grant DEFROST N62909-19-1-2053. Z. P., M. M. and P. S. acknowledge support from projects APVV-20-0425, VEGA 2/0058/20 and VA SR ITMS2014+ 313011W856. G.S.-S. acknowledges financial support from Spanish MINECO Juan de la Cierva program FJCI-2015-25427. Electron microscopy observations were carried out at the Centro Nacional de Microscopia Electronica, CNME-UCM. I. G. and H. S. acknowledge support by the Spanish Research State Agency (FIS2017-84330-R, RYC-2014-15093 and "María de Maeztu" Programme for Units of Excellence in R\&D CEX2018-000805-M), by the Comunidad de Madrid through program NANOMAGCOST-CM (Grant No. S2018/NMT-4321) and SEGAINVEX at UAM. I.G. acknowledges support by the European Research Council PNICTEYES grant agreement 679080.

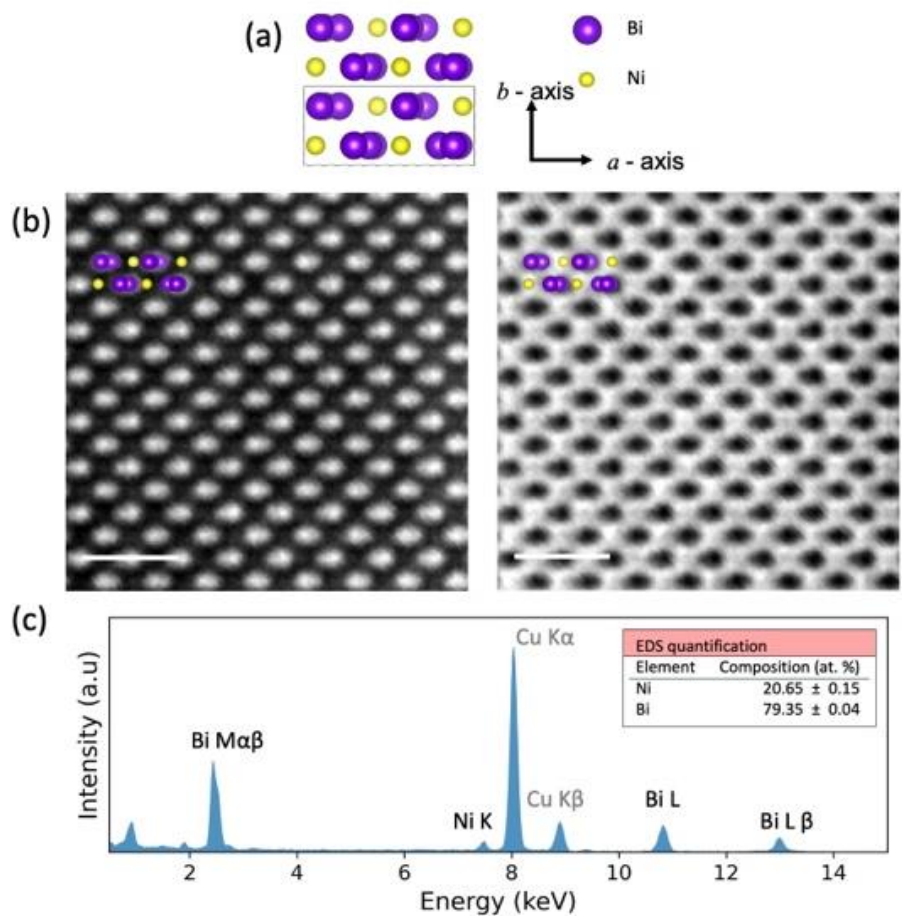

**Figure 1** (a) Schematics of NiBi$_3$ crystalline structure projected in the *a-b* plane. (b) Atomic resolution high angle annular dark field (HAADF) (left panel) and annular bright field (ABF) (right panel) images of NiBi$_3$ in the *a-b* plane. Scale bars correspond to 1nm (c) Averaged EDS spectrum taken at the region shown in (b). Cu peaks originating from the supporting grid are marked in light grey.

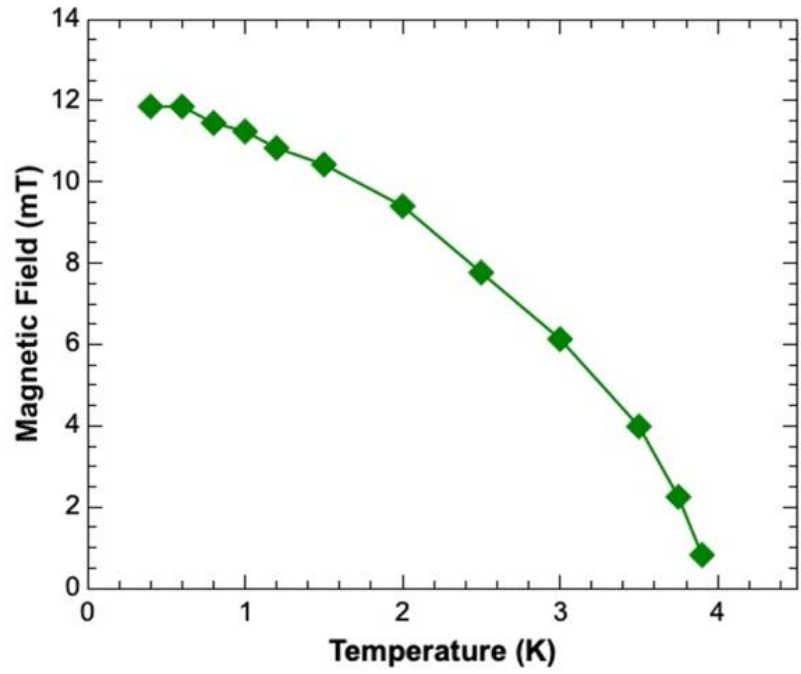

**Figure 2** Lower critical field, $H_{c1}(T)$, obtained from penetration fields measured by local Hall magnetometry at different fixed temperatures.

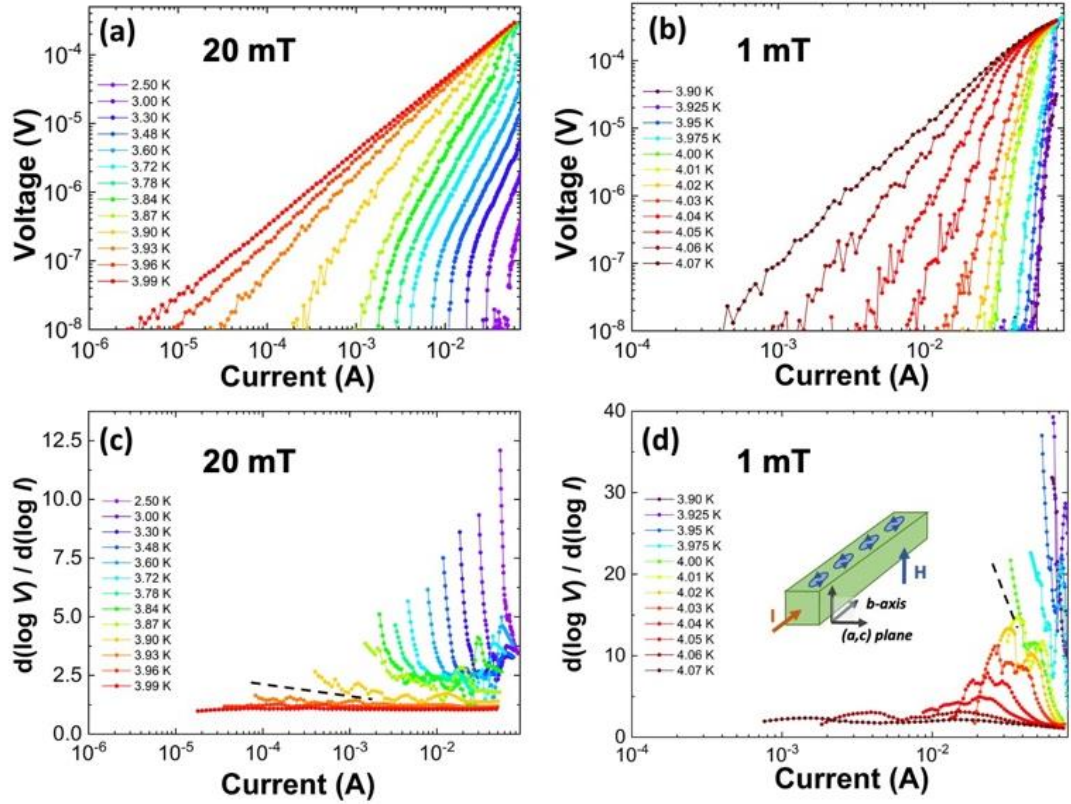

**Figure 3** Voltage vs current (*V-I*) characteristics at different temperatures for (a) 20 mT and (b) 1 mT magnetic fields applied perpendicular to the current direction. (c) and (d) are the derivatives, *d(log V)/d(log I)*, of the curves shown in (a) and (b), respectively. The dashed lines in (c) and (d) indicate the change in the behavior in the derivatives associated with the transition from the vortex liquid to the solid phase. Insert: Sample schematics showing the direction of the current and magnetic field with respect to the crystalline axes of the single crystal (see Fig. 1(a)).

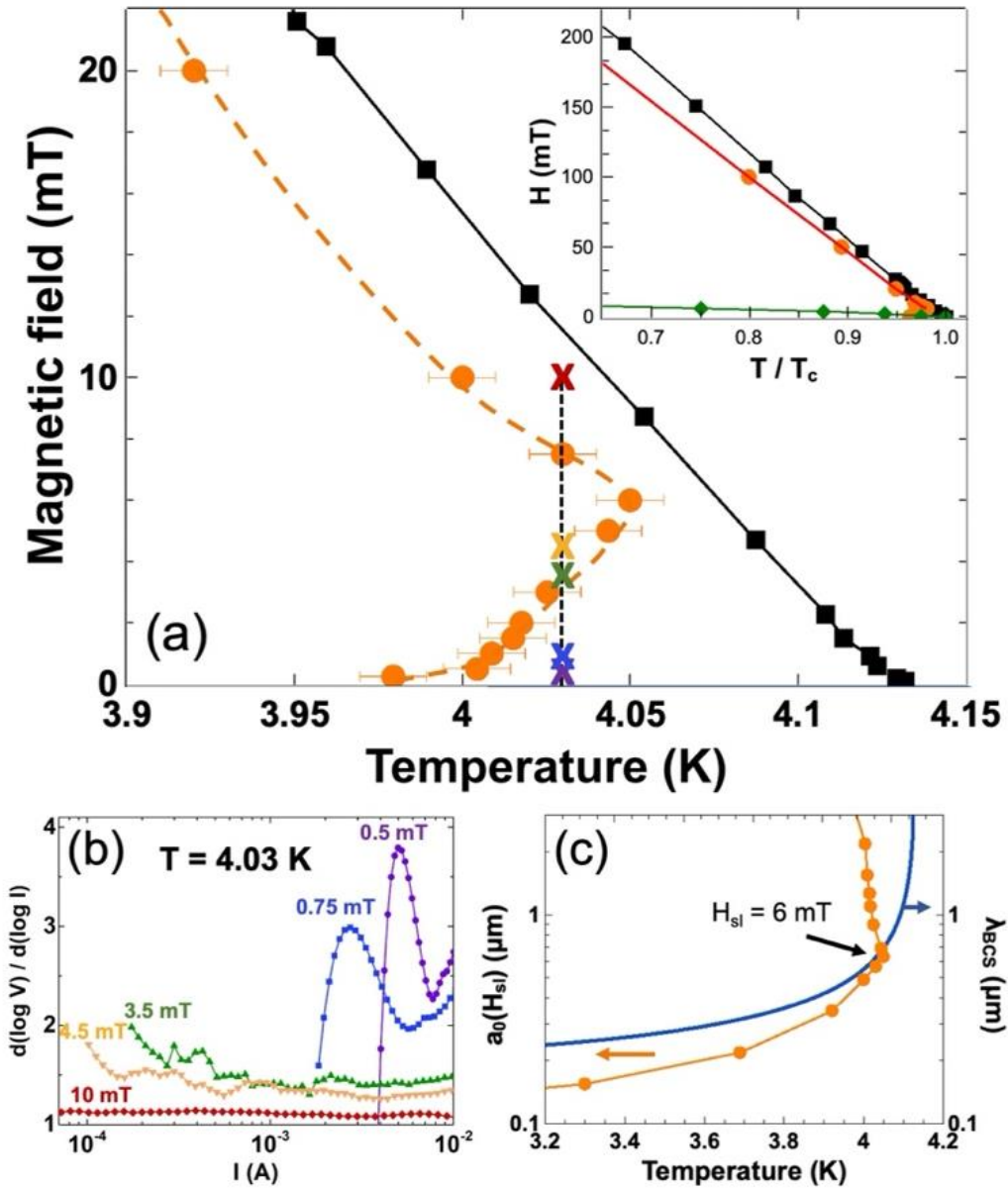

**Figure 4** (a) *H-T* phase diagram for NiBi$_3$ single crystals. The upper critical field, $H_{c2}(T)$ (black squares), is obtained from the resistance vs temperature curves using the 50%R$_N$ criterion as described in the experimental section. The orange circles are obtained from the change in the derivative of *V-I* curves as described in the text indicating the vortex solid-liquid transition (the dashed line is a guide to the eye). Insert: The red line shows the solid-liquid transition line obtained from Eq. (1). The green diamonds correspond to the $H_{c1}$ line from Fig. 2. (b) *d(log V)/d(log I)* vs current at *T* = 4.03 K for different magnetic fields as indicated with coloured crosses in (a). (c) Vortex lattice parameter at the solid-liquid transition, $a_0(H_{sl})$, and $\lambda_{BCS}$ vs temperature. The reentrance of the liquid phase at *H* = 6 mT takes place when $a_0 \sim \lambda$.

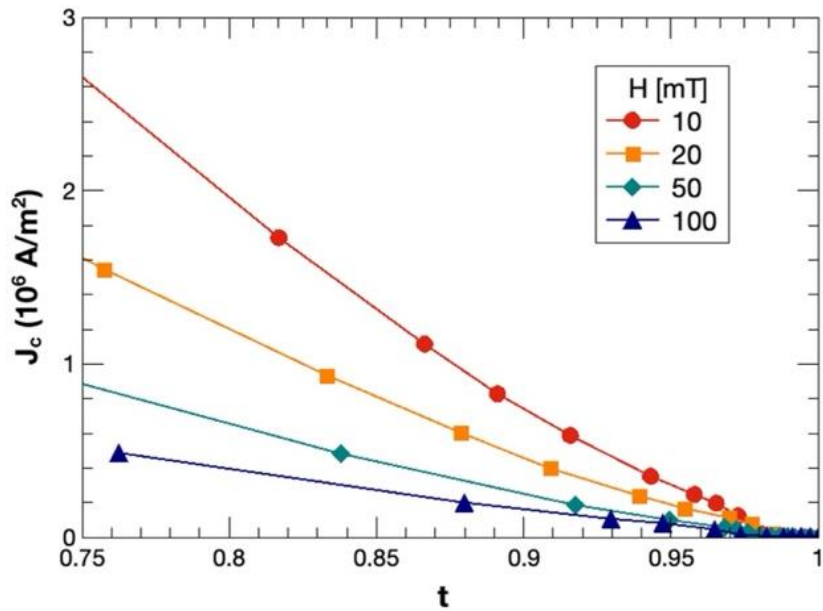

**Figure 5** Critical current as a function of reduced temperature, $t = T/T_{c2}$, for fields higher than 10 mT.

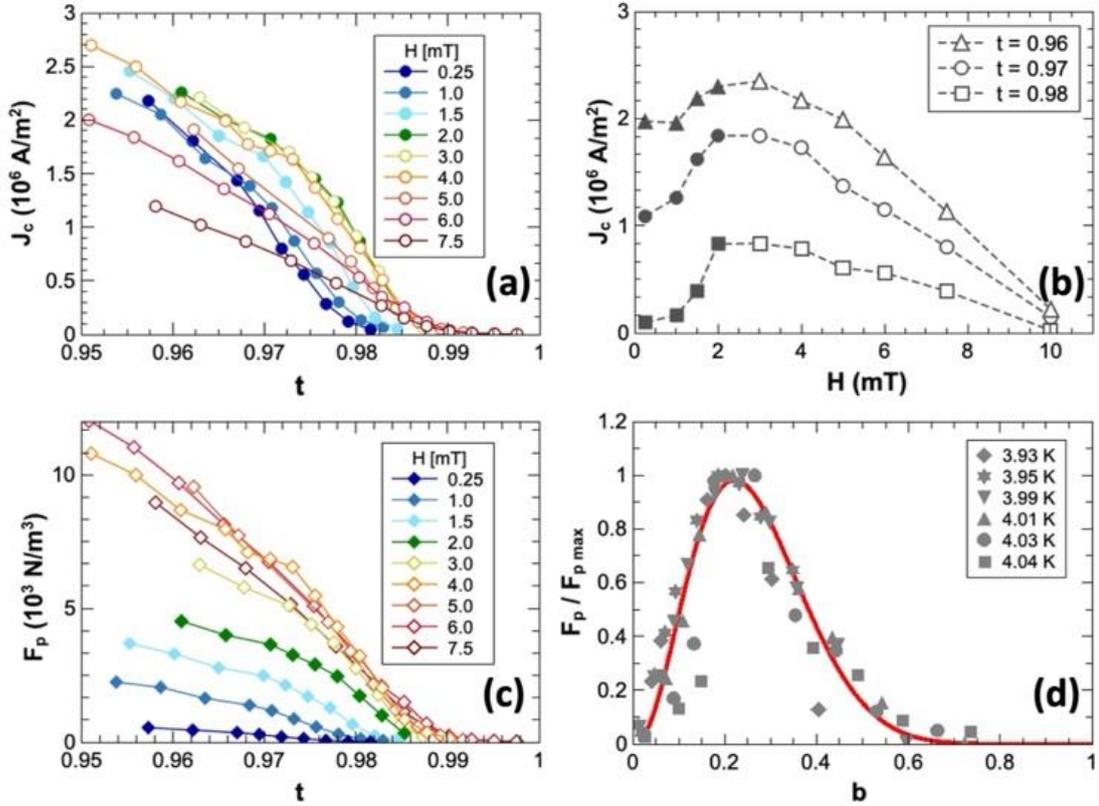

Figure 6 (a) Critical current density as a function of reduced temperature, $t = T/T_{c2}$, for fields below 10 mT. (b) Critical current density vs magnetic field for reduced temperatures $t$ = 0.96, 0.97 and 0.98. The open (filled) symbols correspond to the field range where the critical current increases (decreases) as the magnetic field is reduced. (c) Pinning force calculated from the critical current densities from (a). (d) Normalized pinning force vs reduced field, $b = H/H_{c2}$, in the temperature range between 3.93 K and 4.04 K. The red line corresponds to the relation $F_p \propto b^p(1-b)^q$